# HIGH RESOLUTION HEAVY ION TRACK STRUCTURE IMAGING

G. Laczko*, V. Dangendorf*, M. Krämer+, D. Schardt+, K. Tittelmeier*
* Physikalisch-Technische Bundesanstalt, Braunschweig, Germany
+ Gesellschaft für Schwerionenforschung, Darmstadt, Germany

Short Running Title:
Heavy ion track structure imaging

Address: Gabor Laczko
Physikalisch-Technische Bundesanstalt
P.O.-Box 3345
38023 Braunschweig
Germany
Phone: ++49-531-592-7526
Fax: ++49-531-592-7205
email: glaczko@gmx.de

*Abstract:*

*The difference in the relative biological efficincy (RBE) of ions of same linear energy transfer (LET) but different atomic number (Z) can be attributed to the difference in the radial ionisation distribution. In this contribution we present data from measurements of the spatial ionisation pattern of heavy ions of various Z but similar LET and compare the results with track structure data obtained by Monte Carlo simulations. The measurements were made with a time projection chamber with optical readout (Optical Avalanche Chamber, OPAC) which is able to quantitatively capture the spatial ionisation pattern of an ion traversing the chamber.*

## Introduction

Beams of charged particles like protons or heavier ions represent a promising radiotherapeutic alternative for the treatment of deep-seated tumours. Ion beams have an inverted dose profile: In contrast to the exponential dose deposition of conventional photon beams, the absorbed dose of ions increases with the penetration depth and shows a steep maximum (Bragg peak) and a sharp drop behind it. In addition the RBE of ions is higher than that of sparsely ionising radiation, especially in the stopping region at the Bragg peak. The local effect model (LEM)[1] explains this behaviour by the different distribution of ionisation events. This approach is based on the assumption that the biological effect is entirely determined by the spatial dose distribution in the cell nucleus, and the inactivation probability of a cell is the same due to X-rays and charged particles of the same local dose deposition in the cell nucleus. The dose distribution of ions in a track depends, however on the distance, $r$, from the particle trajectory, roughly following a $1/r^2$ law. The LEM determines the average number of lethal events ($N_{av}$) as :

$$N_{av} = \int_{V_{nucleus}} n(D(r)) \, dr \quad (1)$$

where $n$ is the lethal event density, depending on the local dose, $D(r)$, and $r$ is the position vector. The lethal event density, $n$, is derived from the X-ray survival curve. The main idea of the LEM is not to form the average of the absorbed dose but that of the biological effect. $D(r)$, the spatial energy distribution averaged over many particle tracks, is axially symmetric. Therefore, the main parameter for determining the RBE of a given ion is the LET and the radial dose distribution, $D(r)$, where $r$ is the distance from the track core.

Both, LET and $D(r)$ can be determined by measuring or modelling the transport of ions through matter and sampling the local energy deposition in various biologically relevant volumes. The size of the relevant volumes ranges from a few nm (spacing between DNA helices) to a few μm (cell nucleus). It is not possible to measure with such a high resolution in the condensed phase. Measurements in low pressure gases are therefore performed to model the spatial pattern of ionisation in the condensed phase by a scaling ratio corresponding to the ratio of the densities. The experimental system presented in this paper makes it possible to image the ionisation distribution in a low pressure gas filled chamber with a resolution of down to a few tens of nm in tissue density. The aim of the project is to study the resulting spatial energy deposition distribution of different ions at different energies with emphasis on the radial ionisation distribution ($D(r)$). The experimental results will be compared with results of existing Monte Carlo transport codes[2] which offer another approach to the problem.

## The detector

The experimental method is based on a time projection chamber with a parallel drift field, parallel-plate charge and light amplification layers and optical readout with an image-intensified CCD camera (OPtical Avalanche Chamber, OPAC).

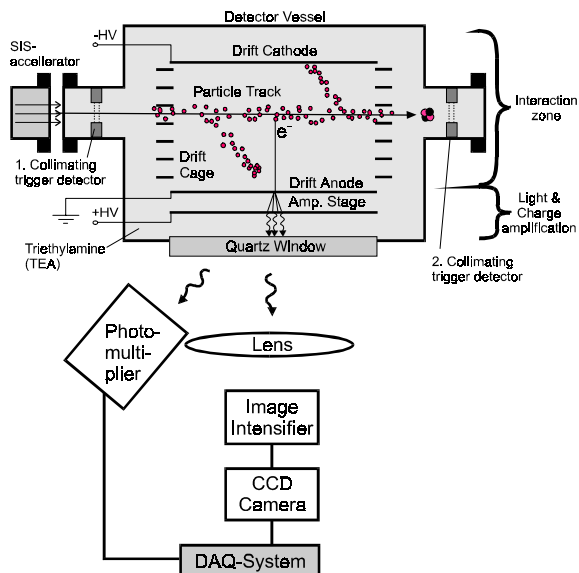

Figure 1. Schematic view of the experimental system

Figure 1 shows the experimental setup as was used in recent experiments with heavy ions at the Gesellschaft für Schwerionenforschung, GSI. The chamber is operated with triethylamine (TEA) vapour in a pressure range of 4 to 40 hPa. By varying the pressure in the chamber it is possible to influence the spatial resolution and the maximum energy of electrons which can be stopped in the chamber. Lower pressure, for example, provides a better spatial resolution when scaled to standard density. The limits of the pressure range are determined by the conditions of stable detector operation. A detailed description of the choice of gas, the electron transport in the gas, the amplification and scintillation properties and the design of the instrument and its optical readout facilities can be found in[3]. Here we will describe only some recent instrumental improvements.

The depth of the drift cage was reduced from 10 cm to 3,2 cm to reduce the diffusion which defines the limit in the spatial resolution of the chamber. The image quality was significantly improved by replacing the image intensifier and the UV-lens. The data acquisition system was upgraded to allow for an image dynamics of 10 bits per pixel (compared to 8 bits before). This is essential to cope with the demand of simultaneous imaging of the intense ion track core and the sparsely ionising parts of δ-electron tracks.

The data acquisition system also records the time-resolved light emission function captured by the photomultiplier tubes and correlates this information with the two-dimensional CCD-image. This feature provides information about the track position in the direction of the electron drift, namely the distance of the track from amplification stage. By means of an improved gating system for the photocathode of the image intensifier, the exposure time of the CCD camera can be varied between 10 ns and a few μs. This feature allows the selection of a narrower subvolume in the drift direction. Another modification was the implementation of two trigger detectors, in front of and behind the interaction volume of OPAC (see figure 1). They are necessary to define the ion trajectory through the detector and generate a signal for the optical read-out electronics.

One of the major problems in the past was the presence of stray light in the images, i.e. not only the direct scintillation light was recorded by the camera but also photons scattered from various detector components and material in the optical path. Numerous efforts (blackening of the detector inside with the low-reflection black paint (Nextel 9218 / 3M)) led, however, to a significant reduction of this artefact.

## Measurement program

Recently, several measurements have been performed with the improved system at the heavy ion synchrotron SIS at GSI. The main focus of the experiments was on carbon ions, because the ion therapy at GSI operates, and the planned therapy facility in Heidelberg is going to operate, with $^{12}$C ions. The aim has been to compare particle tracks of different energies, that are of therapeutic relevance (from 400 MeV/u[*] down to the Bragg peak region). The energies above 100 MeV/u could be provided directly from the accelerator. Below 100 MeV/u, a water column of variable thickness was used to degrade the carbon ion energy. The disadvantage of this method is that at large energy loss in the degrader the energy in OPAC is not very well defined due to energy straggling in the water column. The energy of a particular ion can be determined by measurement of the energy deposition in the sensitive volume of the detector. Three gas pressure values were selected (4, 10 and 40 hPa) to get several views of the ion track structure at different volume scales. About 1000 tracks were recorded at each energy and pressure to obtain reliable statistics even far away from the track core.

Another set of measurements was performed with different kinds of ions as listed in table 1. The guideline for these experiments was to compare data of ions with significantly different Z values, but overlapping LET.

The dependence of the RBE values of different ions at the same LET is described by the local effect model. The track radius (the maximal penetration of δ-electrons from the track core) is proportional to about $E^{1,7}$, where $E$ is the projectile energy. Heavier particles are faster at the same LET, hence they deposit a larger fraction of energy via fast δ-electrons at a greater distance from the particle trajectory. According to the LEM, therefore, integrating and averaging over the biological effect of the local dose around a particle track will generally result in different RBE values for ions of same LET but different Z.

| d$E$/d$x$ (keV/μm) | $^{238}$U (MeV/u) | $^{132}$Xe (MeV/u) | $^{86}$Kr (MeV/u) | $^{36}$Ar (MeV/u) | $^{12}$C (MeV/u) | $^{4}$He (MeV/u) |
|---|---|---|---|---|---|---|
| 70 | | | | 1600 | 28 | 1,8 |
| 275 | | | 1400 | 95 | 5,4 | |
| 680 | | 850 | 110 | 28 | 1,25 | |
| 1840 | 1000 | 117 | 37 | 5,8 | | |
| 3230 | 230 | 51 | 16 | 1,1 | | |

Table 1. Measurement program at GSI for comparison of radial ionisation and cluster size distribution. The stopping power values relate to water, with a density of 1 g/cm$^3$. Different kinds of ions were measured at the same LET (listed in 1st row).

The aim of the experiments is to verify theoretically obtained dose profiles by comparing the measured data with predictions of Monte Carlo (MC) codes. An adaptation of the MC simulation of Krämer (TRAX code[2]) for the geometry and the presence of an electrical field in OPAC has already been developed. The code itself was designed to simulate the emission and transport of δ-electrons after heavy ion impact. Each interaction of the ions or the resulting electrons with the target molecules are treated individually. The type of interaction, the mean free path, the emission angle and energy of the resulting particle are sampled according to the specific cross section. Precise knowledge of the cross sections needed is the most critical factor for reliable MC simulation. In the case of TRAX, experimental data were used as far as possible for the construction of cross-section tables, either directly or via semiempirical fits. However, comprehensive cross-section tables for TEA are not yet available, therefore water vapour, N$_2$ gas or methane must be used in the simulations as target material. In the first approximation, a linear scaling has been applied, depending on the ratio of the number of loosely bounded electrons of TEA and the reference material.

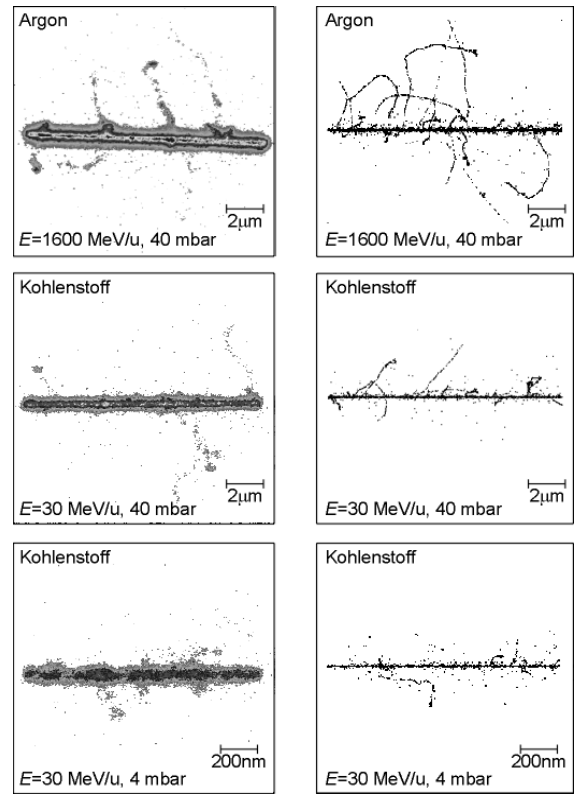

Figure 2:. Some exemplary tracks of the recent measurements (left) and simulations (right). In each cases the LET was 70 keV/μm. The scale bars correspond to distances in condensed matter (ρ = 1 g/cm$^3$). The larger width of the distributions in the measurement is mainly due to electron diffusion.

Some tracks measured with OPAC and corresponding tracks simulated with TRAX are shown in figure 2. It is to be noted that the images are only for demonstration, without statistical relevance. The two upper pairs of images were measured and simulated at 40 hPa gas pressure. The whole length of the tracks is 8 cm in this gas, which corresponds to about 15 μm in tissue. In the third case, 4 hPa pressure was applied: The "zooming factor" is ten times larger. Comparing the two carbon images, it is obvious that the main track is narrower in the 40 hPa case: Most of the δ electrons are stopped very close to the ion trajectory. These images are capable of visualising δ-electrons of higher energies that cannot be stopped in the 4 hPa gas. On the other hand, at 4 hPa, it is possible to analyse the track core in more detail. Referring to table 1, the LET of C and Ar in water are 70 keV/μm in both examples of figure 2, but $^{36}$Ar ions have a 50 times higher energy. Therefore the mean and maximal δ-electron energy is higher for Ar ions, resulting in an energy deposition more expanded in space.

## Guideline for data analysis

Here we report about the main aspects of the data analysis planned as the next step. In the first step the recorded pictures must be cleared from stray light effects which obscure the ionisation density far away from the track core. At high gain, even high energy electrons can be discriminated against stray light events by their quasi-continuous track.

But the data analysis becomes more complicated if the amplification of the chamber is not sufficient to observe this quasi-continuous δ-electron tracks. In the case of $^{238}$U, for instance, the particles typically lose 100 times more energy in the sensitive volume of the detector than carbon ions of the same energy per nucleon. This results in the requirement that the gas amplification has to be dramatically reduced in the case of very high LET particles to ensure stable detector operation without discharges between the grids. The δ-electron tracks around e.g. uranium ions are qualitatively just the same as those of carbon, but occur more frequently. However, due to the reduced gain in the case of U ions, the δ-electron tracks are not amplified as effectively as for the lighter ions displayed in figure 2. For the heaviest ions, only a small sample of ionisation locations of δ-electrons can be observed. Methods are presently developed to distinguish between these sampled δ electrons and the remaining stray light.

The resulting track images are two-dimensional projections of the real track, obscured by different factors such as diffusion of the drifting electrons, exponential amplification functions, optics and data acquisition noise. Comparisons of chosen statistical parameters of the images (e. g. radial ionisation distribution, cluster size distribution) with the corresponding parameters of numerous simulated tracks can be made in two different ways. One method is to unfold the measured images, and the other to fully simulate the detector response. An essential requirement for both is precise knowledge of the response function of OPAC. The diffusion of electrons in homogeneous electric fields has a Gaussian characteristic. The parameters of this transport function are derived from single electron transport experiments. The gain statistics was also measured and can be implemented in the simulation.

## Conclusion and outlook

We have described a particle track chamber with high spatial resolution for applications in radiotherapy and for the general study of ion interactions with matter (e. g. with comic radiation). An overview of the measurement program of summer 2003 at GSI was presented. In future, besides the continuous improvement of the imaging quality, a new strategy will be pursued to reduce the role of electron diffusion. Recent work by Snowden-Ifft et al [4] point out that negative ions (e. g. $CS_2$) can be used as electron carriers. Because of the enormous difference in mass their diffusion is negligible compared to electrons. Moreover, direct imaging of the positive ions from the ionisation processes will be another central effort[5]. Such a device would provide an even better approximation of the radiation effects, because the biologically relevant damage occurs where the positive ions and the electrons are produced and not where the electrons come to rest after being carried away by their initial energy [6]. By means of ion imaging it might be possible to improve the resolution of OPAC down to a few nanometer in tissue.

Footnote:
* The kinetic energy of ions in this paper is quoted in energy per nucleon ($E/u$) with the unit MeV/u